# A narrow-linewidth III-V/Si/Si$_3$N$_4$ laser using multilayer heterogeneous integration

Chao Xiang[1,*], Warren Jin[1], Joel Guo[1], Jonathan D. Peters[1], MJ Kennedy[1], Jennifer Selvidge[2], Paul A. Morton[3], John E. Bowers[1,2]

[1]Department of Electrical and Computer Engineering, University of California, Santa Barbara, Santa Barbara, California 93106, USA

[2]Materials Department, University of California, Santa Barbara, Santa Barbara, California 93106, USA

[3]Morton Photonics, West Friendship, Maryland 21794, USA

*Corresponding author: cxiang@ece.ucsb.edu

**Silicon nitride (Si$_3$N$_4$), as a complementary metal–oxide–semiconductor (CMOS) material, finds wide use in modern integrated circuit (IC) technology. The past decade has witnessed tremendous development of Si$_3$N$_4$ in photonic areas, with innovations in nonlinear photonics[1], optical sensing[2], etc. However, the lack of an integrated laser with high performance prohibits the large-scale integration of Si$_3$N$_4$ waveguides into complex photonic integrated circuits (PICs). Here, we demonstrate a novel III-V/Si/Si$_3$N$_4$ structure to enable efficient electrically pumped lasing in a Si$_3$N$_4$ based laser external cavity. The laser shows superior temperature stability and low phase noise compared with lasers purely dependent on semiconductors. Beyond this, the demonstrated multilayer heterogeneous integration provides a practical path to incorporate efficient optical gain with various low-refractive-index materials. Multilayer heterogeneous integration could extend the capabilities of semiconductor lasers to improve performance and enable a new class of devices such as integrated optical clocks[3] and optical gyroscopes.**

Heterogeneous integration combines the benefits from different material groups which are not natively 'compatible' to offer photonic devices with more functionality[4]. The silicon photonics industry has seen rapid advances since the introduction of heterogeneous III-V integration with silicon to make efficient lasers[5,6]. High-capacity III-V based silicon photonic transceiver chips with seamless integration of lasers and silicon photonic circuits have been commercialized for optical interconnects and millions of devices are being shipped per year[7]. Existing $Si_3N_4$ technology has developed a complete suite of passive devices[8-11] and the next breakthrough requires the addition of an on-chip laser, as active applications still require off-chip laser sources[12-15]. To date, hybrid butt coupling of a III-V gain chip to a $Si_3N_4$ external cavity has been used[16-18]. Compared with a butt-coupled laser, heterogenous integration of $Si_3N_4$ into the laser cavity can eliminate the complicated, expensive optical alignment and packaging of separate chips. Butt coupling also results in much higher coupling loss and parasitic reflection, which are detrimental factors to laser performance.

A heterogeneously fully integrated laser uses compound waveguide coupling through the hybrid evanescent mode within the heterogeneous layers, or via efficient mode transitions between the compound waveguiding layers. In current mature heterogeneous integrated lasers working around 1550 nm, the III-V epitaxial layer has ~ 2 µm thickness and a slab mode refractive index around 3.2, while the refractive index of $Si_3N_4$ is around 2. Even extreme tapering of the thick III-V epitaxial layer is unable to facilitate efficient mode coupling between them within a III-V/$Si_3N_4$ structure. To overcome this, we demonstrate a III-V/Si/$Si_3N_4$ laser structure using multilayer heterogeneous integration that employs multiple mode transitions; from a gain section III-V/Si hybrid waveguide, transitioning to a Si waveguide, through a Si/$Si_3N_4$ transition to the $Si_3N_4$ waveguide. As illustrated in Fig. 1a, the $Si_3N_4$ passive layer is deposited and processed first while the Si layer and subsequent III-V epitaxial layer are transferred on top of the $Si_3N_4$ waveguides via wafer bonding and processed at the backend. The required silicon dioxide ($SiO_2$) cladding for a low loss Si$_3$N$_4$ waveguide consists of a lower thermal oxide cladding and an upper cladding of deposited spacer oxide, together with a VIA oxide layer which is also used for the laser passivation. It is noted that the mature 'SMART CUT' process to make silicon-on-insulator (SOI) wafers can be adopted in the future to fabricate Si-on-Si$_3$N$_4$ wafers[19]. Figure 1b shows a three-dimensional (3D) schematic of the laser. A 1.5 mm long hybrid section with indium phosphide based multiple quantum well (InP MQW) on Si provides the optical gain while the laser mirrors are formed by a narrow-band Si$_3$N$_4$ spiral shaped distributed Bragg grating reflector (DBR) and a broadband tunable Si loop mirror on the two ends. Details of the InP/Si hybrid section design can be found from previous works on InP/Si lasers[20]. Formed by circular Si$_3$N$_4$ posts placed along the curved waveguide (Fig. 1c), the spiral DBR length is 20 mm within a footprint of only 3.5 mm x 3.6 mm using a low loss Si$_3$N$_4$ waveguide (2.8 µm wide and 90 nm thick). The grating period is 526 nm, proving a reflection peak near 1550 nm. The gap between the grating post and waveguide is increased from 920 nm to 944 nm along the spiral to keep the grating unchirped during the decrease of spiral waveguide radius. Another waveguide spiral with its radius decreasing from 100 µm to 40 µm is used as a low-reflection waveguide radiation terminator. With a larger spiral design this terminator could be replaced by a spiral waveguide with opposite curvature to form a laser output for light passing through the grating. The inset picture taken by an infrared (IR) camera shows the Si$_3$N$_4$ spiral grating radiation during lasing. A dual-level Si taper with low reflection is used to adiabatically couple the Si waveguide fundamental transverse electric (TE) mode to the Si$_3$N$_4$ waveguide fundamental TE mode through the hybridized Si/Si$_3$N$_4$ fundamental TE mode in the taper area, over a length of 200 um (Fig. 1d). A thermo-optic tuner is placed between the Si-Si$_3$N$_4$ taper and the InP/Si gain section to enable in-cavity phase tuning. The laser output is taken after the Si loop mirror for characterization. SEM images in Fig. 1e show cross-sectional view of the hybrid InP/Si region, hybrid Si/Si$_3$N$_4$ region and tilted top view of the InP-Si taper.

Figure 2a shows the continuous-wave (CW) light–current–voltage (LIV) characteristics of the III-V/Si/Si$_3$N$_4$ laser operated at a 20 °C stage temperature. The lasing threshold is 75 mA with a peak on-chip output power over 0.5 mW for 320 mA gain current. The differential resistance is about 2.5 Ω. The lasing peak wavelength during the LIV sweep is also recorded using a wavemeter. As the gain current increases, the output power and peak wavelength see several discontinuities following the same trend evidencing a 'cycled' mode hop. This is typical for a DBR laser when gain section temperature increases and the different thermal response time of longitudinal modes and grating reflection peak result in shifted longitudinal mode number[21]. This abrupt wavelength change reflects the longitudinal mode spacing and is about 0.025 nm in our laser. Figure 2b plots four mode states within a continuous wavelength shift cycle when the gain current increases from 232mA to 249 mA. The wavelength red shifts, while power drops as the multimode has less modal gain than the single mode. This mode hop behavior as well as power fluctuation is mainly dependent on the in-cavity optical phase. At 190 mA gain current, the power and peak wavelength show strong hysteresis with phase tuning (Fig. 2c). As the temperature goes down, the laser tends to stay in single-mode operation and remain with relatively high power for a longer time. The Si$_3$N$_4$ spiral grating provides a narrow band filter together with high extinction ratio of sidelobes of over 20 dB. This excellent sidelobe extinction ratio results in a large lasing side mode suppression ratio (SMSR) of over 58 dB (Fig. 2d).

InP lasers and InP/Si lasers are normally quite temperature sensitive as they both have large thermo-optic coefficients. By comparison, the thermo-optic coefficients (dn/dT) of Si$_3$N$_4$ and SiO$_2$ are around 2.45 ×10$^{-5}$ K$^{-1}$ and 9.5x10$^{-6}$ K$^{-1}$ respectively at 1550 nm[22]. This is an order of magnitude smaller than that of Si (dn/dT= 1.8 ×10$^{-4}$ K$^{-1}$) or InP (dn/dT= 2 ×10$^{-4}$ K$^{-1}$)[23,24]. Thus, a laser cavity based on a Si$_3$N$_4$ waveguide DBR will be far less sensitive to temperature variations than one based on Si or InP. We compared our laser with an extended-DBR InP/Si laser with a 15 mm long Si Bragg grating and the results are shown in Fig. 3a[25]. With a temperature change from 10 °C to 55

°C, the wavelength shift of the InP/Si/Si$_3$N$_4$ laser is only 0.47 nm (10.46 pm/ °C), while the InP/Si laser wavelength shift is 3.3 nm (73.18 pm/ °C), over 7x difference.

For many applications such as coherent optical communication and sensing, a narrow spectral linewidth, i.e. a low phase noise is required. The idea of using a long passive cavity to reduce the inherent high phase noise of semiconductor lasers has been extensively studied with III-V/Si heterogeneous lasers using Si waveguide-based cavities[25-27]. Here we measured the frequency noise of our laser (Fig. 3b). With no tuning of the Si loop mirror, it provides ~0.3 power reflectivity. The lowest obtained white-noise-limited frequency noise level is about 2000 Hz$^2$/Hz, giving a Lorentzian linewidth of 6 kHz. This is further reduced to ~ 1300 Hz$^2$/Hz and 4 kHz respectively by tuning the Si loop mirror reflectivity to maximum to increase the photon density in the laser cavity. We also noticed a detuned loading effect in our laser where the minimum linewidth is achieved when the longitudinal mode is slightly red shifted to the spiral grating reflection peak (see Supplementary Information for the detailed measurement)[28]. In this fabrication process run, the spacer oxide thickness is thinner than designed after planarization, resulting in a large mode overlap with the VIA oxide, leading to a propagation loss of 0.43 dB/cm, which can be reduced to around 0.001-0.01 dB/cm in the future[9]. The further reduced loss will enable lasers with a longer Si$_3$N$_4$ spiral grating and weaker coupling constant κ, or cascaded ring resonators, to significantly reduce the laser linewidth and increase the output power[18,29,30]. We anticipate the 320 Hz Lorentzian linewidth achieved with our butt-coupled Si$_3$N$_4$ DBR laser can be further reduced with this new III-V/Si/Si$_3$N$_4$ heterogeneous platform and takes the narrow-linewidth semiconductor lasers to a new level[18].

Low phase noise together with excellent temperature stability can result in ideal lasers for applications where highly stable lasers are required for optical references or sensing. The high temperature stability should also allow this laser to be used in harsh environments at high temperatures. Our approach will also enable a whole new class of devices that require low loss waveguides and are presently not integrated, such as optical gyroscopes which are currently implemented with optical fiber[31], ultra-narrow-linewidth Brillouin lasers and frequency comb generators which require off-chip pump lasers[32]. The successful demonstration of multilayer heterogenous integration is a key step towards fully exploiting the capabilities of wafer bonding technology, which can combine different material features for future multi-functional PICs.

**Methods**

**Fabrication.** The laser fabrication process starts with a 100 mm diameter thermally oxidized crystalline Si wafer. A 90 nm $Si_3N_4$ layer is deposited on top using low pressure chemical vapor deposition (LPCVD). The $Si_3N_4$ waveguide is etched by inductively coupled plasma etching with etching gas $CHF_3/CF_4/O_2$. The LPCVD spacer oxide layer is deposited, annealed and planarized using chemical mechanical polishing (CMP). The remaining thickness of spacer oxide is about 660 nm. A diced SOI wafer with 500 nm thick crystalline Si layer and 1 um thick buried oxide layer is bonded using $O_2$ plasma-assisted direct bonding. The Si substrate removal is mostly done by mechanical polishing and finished by $SF_6/C_4F_8$ based reactive ion etching. The buried oxide layer of bonded SOI wafer is removed by buffered Hydrofluoric acid. Si waveguide is etched by fluorine-based reactive ion etching. An InP epi wafer is cleaved into smaller dies to be bonded to the crystalline Si surface, using $O_2$ plasma-assisted direct bonding. The InP substrate is removed by mechanical polishing and wet etching using diluted Hydrochloric acid. The InP processing then goes though mesa etching though a combination of dry etching using $CH_4/H_2/Ar$ based reactive ion etching and wet etching based on phosphoric acid. The Si layer where it overlaps with $Si_3N_4$ waveguides is selectively etched away except the Si to $Si_3N_4$ taper section before VIA oxide deposition using plasma-enhanced chemical vapor deposition (PECVD). P-contact metal (Pd/Ti/Pd/Au), N-contact metal (Pd/Ge/Pd/Au), heater metal (Ti/Pt) and probe metal (Ti/Au) are deposited using electron-beam evaporation. All the lithography steps are done with a 248 nm deep ultraviolet stepper.

**Mode and taper transmission simulation**. The mode electrical field plot is obtained with COMSOL Multiphysics. Modal index calculation and overlap analysis is performed using Lumerical MODE Solutions. Taper transmission simulation and optimization is done using Lumerical MODE Solutions and FDTD solutions. The detailed taper transmission simulation results are shown in Supplementary Information. In the simulation, the Si layer is 500 nm thick and a shallow-etched Si single-mode ridge waveguide with 250 nm etching depth is used in the InP/Si hybrid section and all Si waveguides except the Si-Si$_3$N$_4$ taper region.

**Frequency noise measurement.** The phase noise measurement is performed using OEWaves phase noise measurement equipment OE4000. An optical fiber isolator with over 40 dB isolation is used after the laser output before the frequency noise characterization.

**Acknowledgments**

The authors acknowledge M. Davenport for useful discussion in design, M. Davenport, C. Zhang, W. Xie, M. Tran and A. Malik for useful discussion in fabrication, and support from Defense Advanced Research Projects Agency (DARPA) (W911NF-19-C-0003).

**Author contributions**

C.X., P.M., J.B. conceived the idea. C.X. led the design, simulation, fabrication, characterization and manuscript preparation. W.J. contributed to the mode simulation. J.G. helped with testing set-up. J.P., M.J., W.J. assisted the fabrication. J.S. took the focused ion beam SEM image. P.M. helped with the grating design. P.M and J.B. participated in the manuscript revision and supervised the project.

**Data availability**

The data that support the findings of this study are available from the corresponding author upon reasonable request.

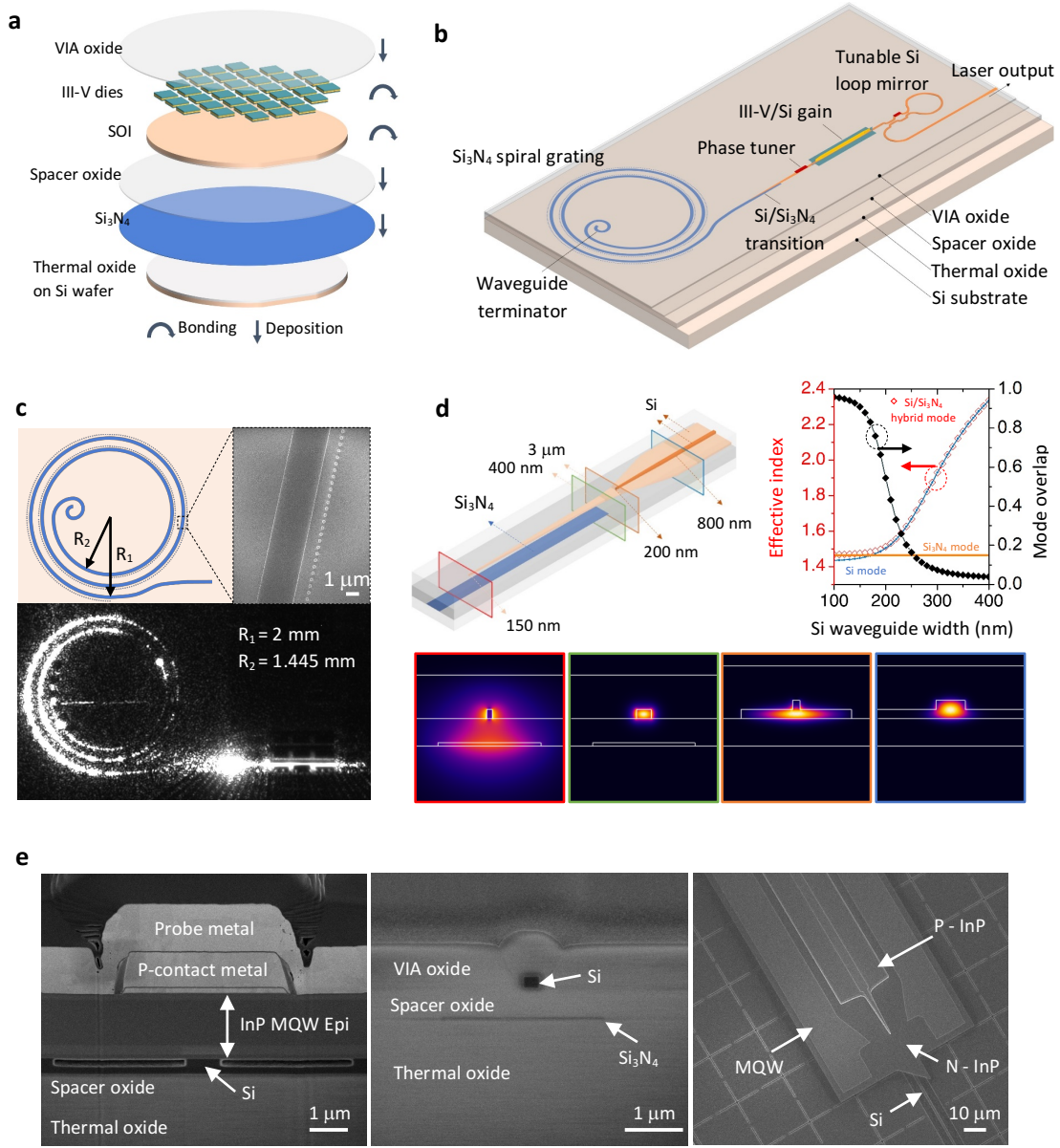

**Figure 1 | Laser schematic design and SEM images.** Schematics of (**a**) multilayer heterogeneous integration, (**b**) III-V/Si/Si$_3$N$_4$ DBR laser based on Si$_3$N$_4$ spiral grating, (**c**) Si$_3$N$_4$ spiral DBR followed by a Si$_3$N$_4$ waveguide terminator, (**d**) Si-Si$_3$N$_4$ taper. Right inset of **c**, SEM image of a section of Si$_3$N$_4$ spiral post grating; bottom inset of **c**, top view IR image taken at lasing with gain current of 150 mA. Right inset of **d**, simulated mode effective index and Si/Si$_3$N$_4$ hybrid mode overlap with Si$_3$N$_4$ mode. Bottom inset of **d**, simulated cross-sectional fundamental TE mode electrical field distribution at different taper section labeled by color contour; red, Si-Si$_3$N$_4$ taper Si$_3$N$_4$ end; green, Si-Si$_3$N$_4$ taper Si start; orange, thick Si to thin Si taper end; blue, thick Si to thin Si taper start. **e**, SEM image of fabricated laser; cross-sectional view at InP/Si (left), Si/Si$_3$N$_4$ (middle) section and tilted top view of InP-Si taper after N-InP mesa formation (right).

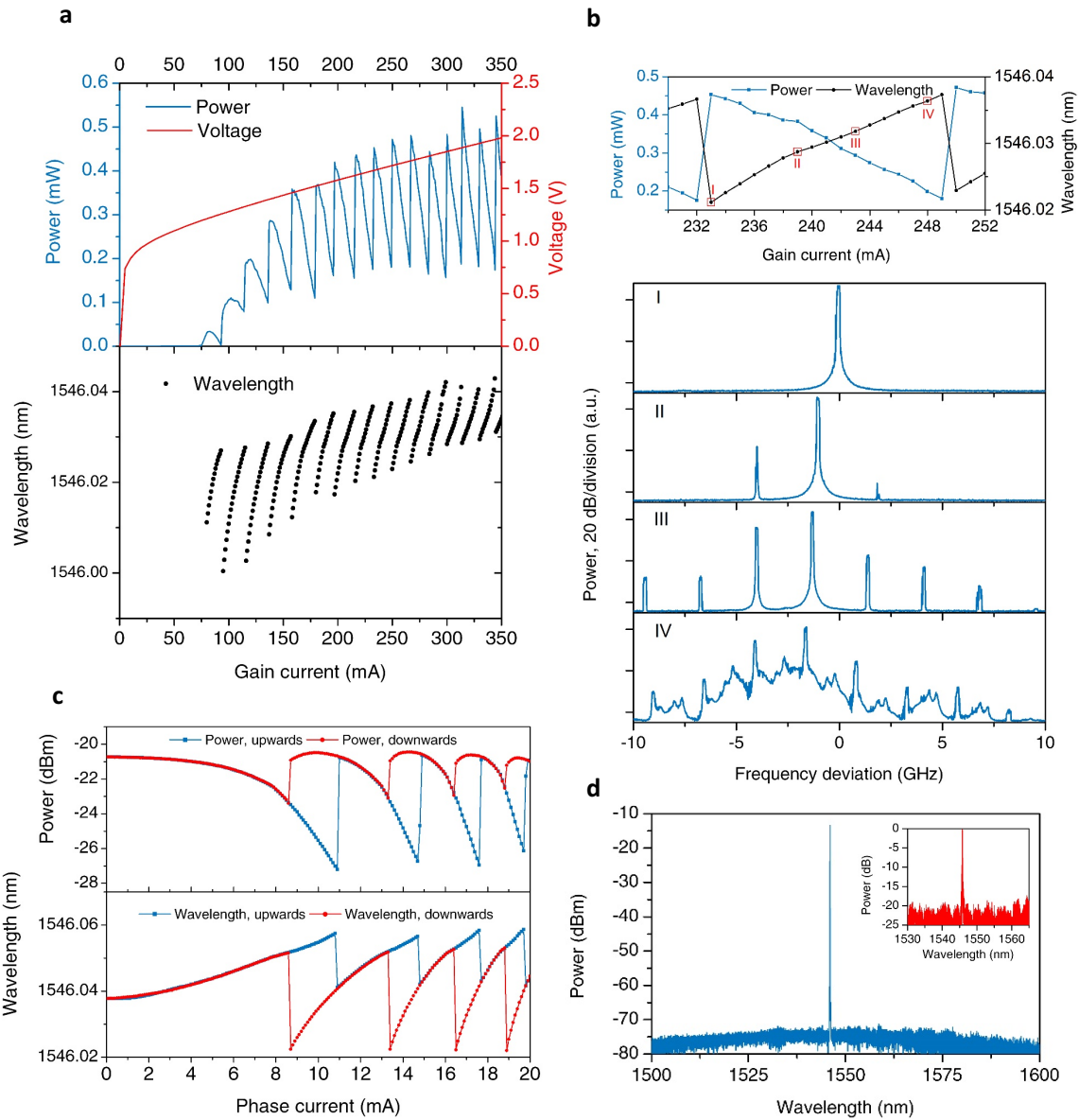

**Figure 2 | Laser performance. a**, LIV characteristics and corresponding peak lasing wavelength. **b**, Close-in lasing spectrum at different bias currents as labeled in the top plot, showing (I) stable single-longitudinal mode, (II, III) stable multi-longitudinal mode and (IV) chaotic multi-longitudinal mode state. **c**, Laser power and peak lasing wavelength hysteresis dependence on the phase current. **d**, Optical spectrum at stable single mode state with gain current of 160 mA. Inset of **d** shows the measured normalized reflection spectra of a passive $Si_3N_4$ spiral grating with identical design used in the laser.

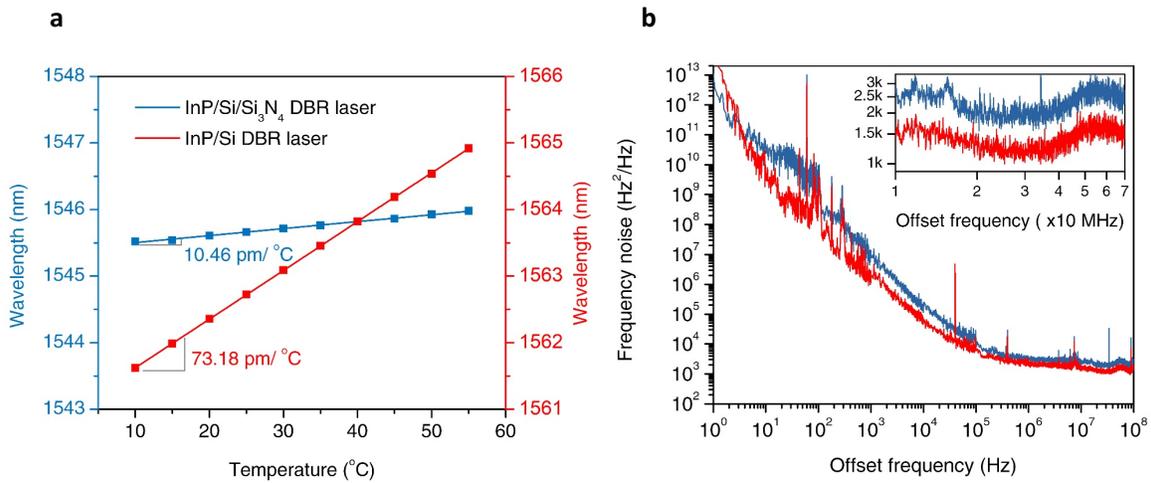

**Figure 3 | Laser temperature stability and frequency noise. a**, Lasing wavelength dependence on stage temperature. Blue, InP/Si/Si$_3$N$_4$ laser at around 250 mA gain current at single mode state; Red, InP/Si DBR laser at around 230 mA gain current at single mode state. **b**, Frequency noise of lasers. Blue, 300.4 mA gain current and zero tuning on Si loop mirror; Red, 320 mA gain current and 10 mA Si loop mirror bias current. Inset of **b**, zoom-in look of white-noise-limited frequency noise level.

**References**

1. Gaeta, A. L., Lipson, M. & Kippenberg, T. J. Photonic-chip-based frequency combs. *Nat. Photonics* **13**, 158–169 (2019).

2. Subramanian, A. Z. *et al.* Silicon and silicon nitride photonic circuits for spectroscopic sensing on-a-chip [Invited]. *Photonics Res.* **3**, B47 (2015).

3. Spencer, D. T. *et al.* An optical-frequency synthesizer using integrated photonics. *Nature* **557**, 81–85 (2018).

4. Komljenovic, T. *et al.* Photonic Integrated Circuits Using Heterogeneous Integration on Silicon. *Proc. IEEE* **106**, 2246–2257 (2018).

5. Fang, A. W. *et al.* Electrically pumped hybrid AlGaInAs-silicon evanescent laser. *Opt. Express* **14**, 9203 (2006).

6. Liang, D. & Bowers, J. E. Recent progress in lasers on silicon. *Nat. Photonics* **4**, 511–517 (2010).

7. Jones, R. *et al.* Heterogeneously Integrated InP/Silicon Photonics: Fabricating Fully Functional Transceivers. *IEEE Nanotechnol. Mag.* **13**, 17–26 (2019).

8. Levy, J. S. *et al.* CMOS-compatible multiple-wavelength oscillator for on-chip optical interconnects. *Nat. Photonics* **4**, 37–40 (2010).

9. Bauters, J. F. *et al.* Planar waveguides with less than 0.1 dB/m propagation loss fabricated with wafer bonding. *Opt. Express* **19**, 24090 (2011).

10. Spencer, D. T. *et al.* Low kappa, narrow bandwidth $Si_3N_4$ Bragg gratings. *Opt. Express* **23**, 30329 (2015).

11. Xiang, C., Davenport, M. L., Khurgin, J. B., Morton, P. A. & Bowers, J. E. Low-Loss Continuously Tunable Optical True Time Delay Based on $Si_3N_4$ Ring Resonators. *IEEE J. Sel. Top. Quantum Electron.* **24**, 1–9 (2018).

12. Blumenthal, D. J., Heideman, R., Geuzebroek, D., Leinse, A. & Roeloffzen, C. Silicon Nitride in Silicon Photonics. *Proc. IEEE* **106**, 2209–2231 (2018).

13. Gundavarapu, S. *et al.* Sub-hertz fundamental linewidth photonic integrated Brillouin laser. *Nat. Photonics* **13**, 60–67 (2019).

14. Xie, W. *et al.* On-Chip Integrated Quantum-Dot-Silicon-Nitride Microdisk Lasers. *Adv. Mater.* **29**, 1604866 (2017).

15. Sacher, W. D. *et al.* Monolithically Integrated Multilayer Silicon Nitride-on-Silicon Waveguide Platforms for 3-D Photonic Circuits and Devices. *Proc. IEEE* **106**, 2232–2245 (2018).

16. Fan, Y. et al. 290 Hz intrinsic linewidth from an integrated optical chip based widely tunable InP-$Si_3N_4$ hybrid laser. In *Proc. CLEO* Paper JTh5C.9 (OSA, 2017).

17. Stern, B., Ji, X., Dutt, A. & Lipson, M. Compact narrow-linewidth integrated laser based on

a low-loss silicon nitride ring resonator. *Opt. Lett.* **42**, 4541 (2017).

18. Xiang, C., Morton, P. A. & Bowers, J. E. Ultra-narrow linewidth laser based on a semiconductor gain chip and extended $Si_3N_4$ Bragg grating. *Opt. Lett.* **44**, 3825 (2019).

19. Bruel, M., Aspar, B. & Auberton-Hervé, A.-J. Smart-Cut: A New Silicon On Insulator Material Technology Based on Hydrogen Implantation and Wafer Bonding. *Jpn. J. Appl. Phys.* **36**, 1636–1641 (1997).

20. Davenport, M. L., Liu, S. & Bowers, J. E. Integrated heterogeneous silicon/III–V mode-locked lasers. *Photonics Res.* **6**, 468 (2018).

21. Coldren, L. A., Corzine, S. W. & Mašanović, M. L. *Diode Lasers and Photonic Integrated Circuits*. (John Wiley & Sons, Inc., 2012). doi:10.1002/9781118148167

22. Arbabi, A. & Goddard, L. L. Measurements of the refractive indices and thermo-optic coefficients of $Si_3N_4$ and $SiO_x$ using microring resonances. *Opt. Lett.* **38**, 3878 (2013).

23. Cocorullo, G. & Rendina, I. Thermo-optical modulation at 1.5µm in silicon etalon. *Electron. Lett.* **28**, 83–85 (1992).

24. Martin, P., Skouri, E. M., Chusseau, L., Alibert, C. & Bissessur, H. Accurate refractive index measurements of doped and undoped InP by a grating coupling technique. *Appl. Phys. Lett.* **67**, 881–883 (1995).

25. Huang, D. *et al.* High-power sub-kHz linewidth lasers fully integrated on silicon. *Optica* **6**, 745 (2019).

26. Kazarinov, R. & Henry, C. The relation of line narrowing and chirp reduction resulting from the coupling of a semiconductor laser to passive resonator. *IEEE J. Quantum Electron.* **23**, 1401–1409 (1987).

27. Tran, M. A. *et al.* Ring-Resonator Based Widely-Tunable Narrow-Linewidth Si/InP Integrated Lasers. *IEEE J. Sel. Top. Quantum Electron.* **26**, 1–14 (2020).

28. Vahala, K. & Yariv, A. Detuned loading in coupled cavity semiconductor lasers—effect on quantum noise and dynamics. *Appl. Phys. Lett.* **45**, 501–503 (1984).

29. Xiang, C., Morton, P. A., Khurgin, J., Morton, C. & Bowers, J. E. Widely Tunable $Si_3N_4$ Triple-Ring and Quad-Ring Resonator Laser Reflectors and Filters. in *Proc. IEEE 15th Int Conf. Group IV Photon.* (IEEE, 2018).

30. Morton, P. A. & Morton, M. J. High-Power, Ultra-Low Noise Hybrid Lasers for Microwave Photonics and Optical Sensing. *J. Light. Technol.* **36**, 5048–5057 (2018).

31. Bergh, R., Lefevre, H. & Shaw, H. An overview of fiber-optic gyroscopes. *J. Light. Technol.* **2**, 91–107 (1984).

32. Stern, B., Ji, X., Okawachi, Y., Gaeta, A. L. & Lipson, M. Battery-operated integrated frequency comb generator. *Nature* **562**, 401–405 (2018).